\documentclass{article}
\usepackage{spconf}
\usepackage{subcaption}
\usepackage{multirow}

\usepackage{cite}
\usepackage{amsmath,amssymb,amsfonts}
\usepackage{algorithm}
\usepackage{algpseudocode}
\usepackage{amsmath}
\usepackage{graphics}
\usepackage{epsfig}
\usepackage{graphicx}
\usepackage{textcomp}
\usepackage{xcolor}
\usepackage[absolute]{textpos}

\algtext*{EndWhile}
\algtext*{EndIf}
\algtext*{EndFor}

\title{A Study on speech enhancement using exponent-only floating point quantized neural network (EOFP-QNN)}
\name{Yi-Te Hsu$^{1}$, Yu-Chen Lin$^{2}$, Szu-Wei Fu$^{1,2}$, Yu Tsao$^{1}$, Tei-Wei Kuo$^{1,2}$}
\address{
$^{1}$Research Center for Information Technology Innovation, Academia Sinica, Taiwan\\
$^{2}$Department of Computer Science and Information Engineering, National Taiwan University\\
{\small b01901112@ntu.edu.tw, f04922077@csie.ntu.edu.tw, d04922007@ntu.edu.tw, yu.tsao@citi.sinica.edu.tw, ktw@csie.ntu.edu.tw}
}
\begin{document}


%
\maketitle
\begin{abstract}


Numerous studies have investigated the effectiveness of neural network quantization on pattern classification tasks. The present study, for the first time, investigated the performance of speech enhancement (a regression task in speech processing) using a novel exponent-only floating-point quantized neural network (EOFP-QNN). The proposed EOFP-QNN consists of two stages: mantissa-quantization and exponent-quantization. In the mantissa-quantization stage, EOFP-QNN learns how to quantize the mantissa bits of the model parameters while preserving the regression accuracy in the least mantissa precision. In the exponent-quantization stage, the exponent part of the parameters is further quantized without any additional performance degradation. We evaluated the proposed EOFP quantization technique on two types of neural networks, namely, bidirectional long short-term memory (BLSTM) and fully convolutional neural network (FCN), on a speech enhancement task. Experimental results showed that the model sizes can be significantly reduced (the model sizes of the quantized BLSTM and FCN models were only 18.75\% and 21.89\%, respectively, compared to those of the original models) while maintaining a satisfactory speech-enhancement performance.


\end{abstract}
\begin{keywords}
Speech Enhancement, Quantized Neural Networks, Floating Point, Embedded Devices
\end{keywords}

\section{Introduction}\label{sec:Introduction}
In the past few years, deep learning (DL)-based models have been widely used in many different applications. Because of their deep structures, DL-based models can effectively extract representative features when performing classification and regression tasks. It has been confirmed that DL-based approaches outperform traditional methods in image recognition \cite{imgRec1, ResNet2016}, speech recognition \cite{Speech2013,Speech2012,Jinyu2013}, object detection \cite{obj1,obj2,obj3}, and natural language processing \cite{refNLU, Mikolov2013, Mikolov2012}.
On the other hand, also because of their deep structures, DL-based approaches generally require larger storage and higher computational costs than traditional methods. 
To meet these demands, many physical hardware devices, such as graphics processing units and tensor processing units \cite{refTPU}, have been developed.
Furthermore, to facilitate real-time predictions in an Internet-of-Things (IoT)\cite{refIoT} system, researchers also seek solutions to install DL-based models in embedded systems. One potential solution to this goal is to compress DL-based models by using some quantization technique. 

Numerous model quantization techniques have been proposed.
Courbariaux et al. \cite{BinaryConnect} proposed the BinaryConnect algorithm, which uses only 1 bit for all the weights in the model. 
Experimental results showed that the quantized model still yielded state-of-the-art classification results.
Gong et al. \cite{G2014} used a clustering technique to find the respective floating-point centroid values to replace the original weights. 
Experimental results showed that the compressed model can still obtain 1\% loss in the top-5 recognition rate.
Zhou et al. \cite{INQ} proposed an incremental network quantization (INQ) technique to convert a pre-trained full-precision convolutional neural network (CNN) model into a low-precision one. Based on this technique, all the weights are constrained to be powers of 2 and 0. In the end, INQ uses 5 bit-width to reach slightly better results in terms of top-1 and top-5 errors. 
Based on our literature survey, most compression techniques are proposed to be applied to DL-based models for classification tasks, such as image and speech recognition\cite{G2014,INQ,bridgeC,qspeech1,qspeech2,qspeech3,qspeech4,qspeech5,qspeech6}.
Unlike the output of a classification task, which classifies the data into a set of categories, the output of a regression task will be continuous variables. Owing to the different output formats, the effect of model compression on regression tasks should be very different from that on classification tasks \cite{ko2017precision}. In this study, we focus our attention on deriving a new model quantization algorithm for the speech enhancement (SE) task, which is an important regression task in speech signal processing. 

The goal of DL-based SE is to map a noisy speech signal into an enhanced one with improved intelligibility and quality\cite{x1}. 
One class of DL-based SE methods performs spectral mapping that aims to transform noisy spectral features to clean ones. 
Effective models used to characterize the mapping function include deep denoising autoencoder (DDAE)\cite{x2,xia2014wiener}, deep neural network (DNN)\cite{x3,kolbk2017speech}, CNN \cite{x4,fu2017complex}, and bidirectional long short-term memory (BLSTM)\cite{x5,chen2015speech}. 
Another class of approaches aims to map a noise speech waveform to a clean one directly to address possible distortions caused by imperfect phase information in the spectral mapping-based approaches, and the fully convolutional network (FCN) is usually used to characterize the mapping function \cite{fu2017raw, x6, x7, x8}. Clearly, the storage and computational cost are important factors to be considered when implementing SE models in practical applications. However, based on our literature survey, only few studies have investigated potential approaches to reduce the model complexity of SE models. In \cite{sun2017optimization}, a weights sharing and quantization technique is used to compress a DNN-based SE model. The results showed that although the model size can be reduced, the enhanced speech quality was notably reduced as compared to the one generated by the original model. Meanwhile, Ko et al. investigates the relation of precision scaling process and SE performance  \cite{ko2017precision}. The results showed that removing too many bits can cause notable performance degradation. For practical applications, the storage requirement and achievable performance need to be considered simultaneously. To strike a good balance of storage requirement and performance, this study proposes a novel exponent-only floating-point (EOFP) neural network quantization technique for the SE task. We selected one spectral mapping approach, namely, BLSTM, and one waveform mapping approach, namely, FCN, to test the EOFP-QNN on an SE task.   



The remainder of this paper is organized as follows.
Section~\ref{sec:background} introduces the related works and the motivation for the study.
Section~\ref{sec:design} details the proposed EOFP-QNN.
Section~\ref{sec:Experiments} presents the experimental setup and results.
Finally, Section~\ref{sec:conclusion} provides the concluding remarks.
\section{Background and Motivation}\label{sec:background}
\subsection{Speech enhancement}

 
Generally speaking, a DL-based SE includes two phases: offline and online. In the offline phase, the noisy speech is first fed into the SE system to generate the enhanced speech. A criterion is used to evaluate the difference of the enhanced speech and the clean references, such as the mean square error (MSE)\cite{x2,x3,x4,x5}, L1 norm\cite{x7}, or short-time objective intelligibility measure (STOI)\cite{x6}. Then, the difference is used to update the parameters in the DL-based models. In the online phase, the noisy signal is fed into the trained SE system to obtain the enhanced speech.

Conventionally, SE is applied on noisy speech with time-frequency representation (e.g., magnitude spectrum). Thus, an additional signal analysis process is needed to convert speech waveforms to spectral features before denoising. To reconstruct the waveform domain, most of spectral mapping-based methods simply borrow the phase from the noisy speech. Recently, waveform mapping-based methods \cite{x6} are also proposed, which directly enhance the noisy waveform. 
In this study, we selected BLSTM and FCN for the spectral mapping and waveform mapping, respectively, as two representatives to the regression tasks. The parameters in the BLSTM and FCN models before compression were 32-bit floating-point values, and their performances were investigated after quantizing to the proposed EOFP-QNNs.

\subsection{Floating-Point Representation in a DL-based Model}

\begin{figure}[ht]
\centering
\includegraphics[width=1\columnwidth]{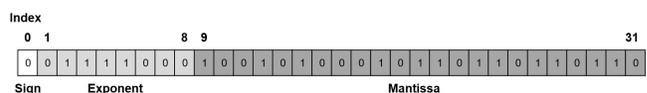}
\caption{Single precision floating point format.}
\label{fig:binary32}
\end{figure}

A DL-based model generally has a large number of parameters, and most of these parameters are stored in a floating-point data format. The single-precision floating-point format of IEEE 754 \cite{IEEE754} is the most common format to represent the parameters, and its binary format is shown in Fig.~\ref{fig:binary32}. In the figure, a single-precision floating-point value, $bits$ consists of three parts: $bits[0]$ indicates the \emph{sign}, $bits[1:8]$ represent the \emph{exponent} (an unsigned integer), and $bits[9:31]$ stand for the \emph{mantissa} (or significand or fraction). Except for the sign, the exponent, and the mantissa are not represented directly. Because the exponent is an unsigned integer, the smallest representable exponent must be shifted to 1 by an additional \emph{bias}. Thus, the decimal value of single-precision data can be calculated by the following equation:
\begin{equation}
(value)_{10} = (-1)^{sign} * (mantissa)_{10} * 2^{(exponent)_{10} - bias}
\end{equation}
where the bias is 127 ($2^{7}-1$); the decimal mantissa value is
\begin{equation}
(mantissa)_{10} = 1 + \sum_{i = 1}^{23} b_{8+i}*2^{-i}
\label{eq:mantissa-equation}
\end{equation}
Taking Fig.~\ref{fig:binary32} for example, we obtain the decimal value of the single-precision data as 0.012339999\ldots.

Now we can see that \emph{sign} and \emph{exponent} determine the range, whereas \emph{mantissa} determines the precision. 
Some applications may need high-precision floating points, while such floating points are not always necessary.
Table~\ref{tab:motivation} shows a preliminary experiment based on a BLSTM SE model. For all the parameters in a neural network model, we masked a sequence of bits to zero from the end. 
As a result, the decimal values of parameters were changed, yet the enhanced speech presented similar qualities. 
The preliminary experiment result in Table~\ref{tab:motivation} inspired us to quantize the floating-point data parameters and thus compress the SE models.
\begin{table}[h]
	\caption{A preliminary experiments on speech quality and intelligibility under different precisions.} 
	\centering 
	\begin{tabular}{c c c c c }
		\hline\hline
		Mask & \multirow{2}{*}{Binary} & \multirow{2}{*}{Decimal} & \multirow{2}{*}{PESQ}\\
		Bits & & & & \\
		0 & 0\ldots110110110110 & 0.012339999\ldots & 2.144\\ 
		6 & 0\ldots110110\textbf{000000} & 0.012339949\ldots & 2.135\\ 
		12 & 0\ldots\textbf{000000000000} & 0.012336730\ldots & 2.141\\ 
		\hline 
	\end{tabular}
	\label{tab:motivation}
\end{table}

\begin{figure*}[ht]
	\centering
	\includegraphics[height=4.8cm]{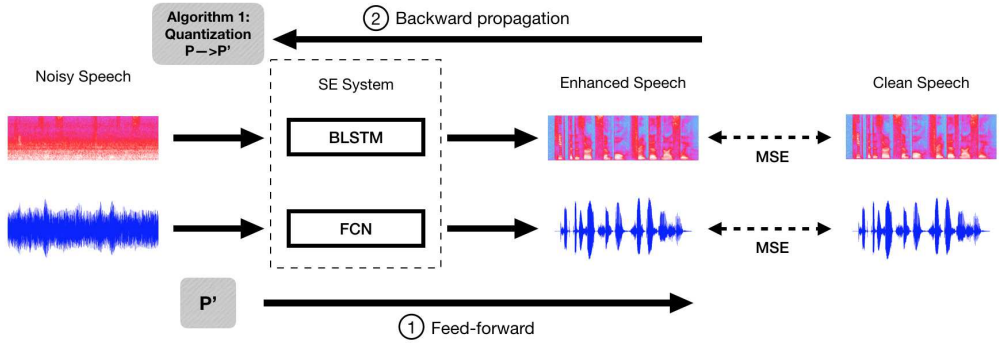}
	\caption{The training procedure of the mantissa-quantization.}
	\label{fig:network}
\end{figure*}
Nevertheless, two setups need to be determined in the proposed algorithm. The first one is to decide when the quantization should be executed. 
The second one is to determine the appropriate number of bits to quantize, to ensure that the enhanced speech have similar qualities to that with the original single-precision value.

\section{EOFP-QNN}
\label{sec:design}

This section presents the proposed EOFP quantization technique for SE models. First, we introduce the overall procedure of the EOFP-QNN. Then, we detail the philosophies of the mantissa-quantization and exponent-quantization.

\subsection{Overall Procedure of the Model Quantization}
An intuitive quantization method is to load out the model (the parameters are saved in a single-precision format) after training and directly quantize the parameters in the model for testing. Because the parameters are trained based on the original single-precision format, this direct quantization approach may cause performance degradations.
To overcome this issue, we intend to make the model learn how to quantize the parameters during the training phase.
Fig.~\ref{fig:network} shows the procedure of the SE neural network model with our quantization and parameter update, in two gray boxes.

As shown in Fig.~\ref{fig:network}, we execute the quantization at the end of each epoch. More specifically, we allow the model to train, to learn the precise weights in single-precision parameters within one epoch. After one epoch is completed, we quantize all the parameters and force them to be less bit-width parameters. Note that the bit-width denotes the number of remaining bits after quantization. Then, the model must use the quantized parameters in the feed-forward part in the following epoch. At the end of training, we get the quantized SE model. In short, we adopt a straightforward solution that directly quantizes the parameters \textbf{P} in each epoch. The quantized parameters \textbf{P'} will be fed in the next training epoch.

\subsection{Mantissa-quantization}
The percentage of the mantissa part is $\frac{23}{32}$, which approaches 72\% in the single-precision floating point.
Therefore, we choose to quantize the parameters with less-precision mantissa as our first step.
Before quantization, we first define the target number $n$ of bits to quantize because different applications may be tolerant to different precision. The natural limitation of $n$ is in the range [0:23] since only 23 bits are used in the mantissa part. It is noted that $n=0$ means that no quantization process is applied.
\begin{algorithm}[h]
	\caption{Mantissa-quantization}
	\begin{algorithmic}[1]
		\Require
		The target number of bits $n$ to quantize, for a model with $l$ layers, $\{L_i|i=1, 2, \ldots, l\}$
		\State $mask \gets [1_{0}1_{1}\ldots1_{31-n}0_{31-n+1}0_{31-n+2}\ldots0_{31}]$
		\For{each layer $L_i$}
		\For {each floating point parameter $p$ in $L_i$}
			\State Convert $p$ to 32-bit binary $bits[0:31]$
			\If {$23>n>0$}
				\State $bits[31-n] = bits[31-n]$ \textbf{\(\vert \vert\)} $bits[31-n+1]$
			\ElsIf {$n=23$}
				\State $bits[1:8] = bits[1:8] ~\textcircled{+} ~ bits[9]$ ~,~where \textcircled{+} is the binary addition operator
			\EndIf
			\State $bits[0:31] = bits[0:31]$ \& $mask[0:31]$
			\State Convert $bits$ back to $p$
	\EndFor
	\EndFor
	\end{algorithmic}
	\label{algo:mantissa}
\end{algorithm}

Algorithm~\ref{algo:mantissa} presents the mantissa-quantization, which is executed after the backward-propagation.
More specifically, after the backward propagation, Algorithm~\ref{algo:mantissa} first defines a 32-bit-length binary $mask$, where the head $32 - n$ bits are 1’s and the latter $n$ bits are 0’s.
For each parameter $p$ in the layer of the model, a conditional rounding arithmetic is used to quantize the value of the mantissa part.
We first convert the data type of $p$ from floating point to binary format $bits$.
If the target number $n$ is greater than 0 and less than 23 (i.e., $23 - n$ bits are remaining in the mantissa part), then the value of $bits[31 - n]$ is obtained by performing the OR operation on original $bits[31 - n]$ and $bits[32 - n]$.
If $n$ is $23$, meaning there is no bit left in the mantissa part, then the whole exponent partition, $bits[1 : 8]$, is added to the value of $bits[9]$, the first bit of the mantissa.
The main reason that we divided our algorithm into two cases, $23 > n > 0$ and $n = 23$, was to avoid an overflow problem in the mantissa part. As mentioned in Section 2.2, the exponent represents an unsigned integer, and thus it is not possible that 
$bits [1 : 8]$ are all 1s.
Thus, we can directly use the rounding arithmetic to quantize.
However, it is possible that 
$bits [9 : 31-n]$ are all 1s in the mantissa part of all parameters
that may cause an overflow problem. Accordingly, we propose our conditional rounding arithmetic to calculate the value of the last bit only. 
The last $n$ bits are removed by taking the intersection with the binary $mask$, and the binary $bits$ is converted back to the floating point $p$. 
After all the parameters are updated (quantized), the feed-forward process is then performed using the quantized model.




\subsection{Exponent-quantization}
According to the format of the single-precision floating point, it is obvious that there are at most 23 bits that we can quantize in the mantissa-quantization, where there are 9 bits remaining.
As mentioned in Section 2.2, the single-precision floating-point format provides a \emph{bias} by helping the exponent part represent the range from  $2^{-127}$ to $2^{128}$, where $2^{-127}$ and $2^{128}$ are defined as 0 and $\infty$, respectively.
Taking the sign bit into consideration, we have the exact range of the remaining 9 bits as $\pm0, \pm2^{-126}$ to $\pm2^{127}$, and $\pm\infty$, which is a great range to represent values in general.

However, the normalization process, which is often executed during the DL-based model training, restricts the value of the parameter in a certain range, and, thus, there are only marginal differences among the values of the parameters in each layer. 
In other words, after the normalization, it is not necessary to represent the parameters using a wide range of floating point format.
Therefore, we propose the statistical exponent-quantization to further compress the DL-based model by analyzing the distribution of the parameter values.


\begin{algorithm}[h]
	\caption{Exponent-quantization}
	\begin{algorithmic}[1]
		\Require The neural network model $\Lambda$
		\Ensure  the length $len$, minimum exponent log$_{2}$ value $min$, a  quantized model $\Lambda'$
		\State Find the maximum and minimum exponent log$_{2}$ value, $max$ and $min$ of all parameters, except for zero value
		\State $len$ = $Ceil$\{log$_{2}[(max-min+1)+1]$\}
		\For {each parameter $p$ in $\Lambda$}
			\State Fetch the exponent $(e)_{10}$ of $p$
			\If{$p = 0$}
				\State $(e')_{10} = 0$
			\Else
				\State $(e')_{10} = (e)_{10} - min+1$
			\EndIf
			\State $p'$ with $(e')_{10}$ as exponent is replaced in $\Lambda'$
		\EndFor

		\State \Return $len, min, \Lambda'$
%
	\end{algorithmic}
	\label{algo:post}
\end{algorithm}

Algorithm~\ref{algo:post} presents the exponent-quantization for each parameter. The output of the quantized parameter includes three attributes.
Before quantizing, we need to calculate the least number of bits $len$ that can represent the range of all parameters.
We first determine the maximum and minimum log$_{2}$ values, except for the zero value, among all parameters in the model $\Lambda$.
Then, we calculate the least length $len$ by applying the ceiling function to log$_{2}[(max-min+1)+1]$. 
The last 1 in the equation is to represent one more value, zero, which cannot be written in the power of 2.
In the exponent-quantization, we first fetch the exponent part $(e)_{10}$ of $p$ for every parameter in $\Lambda$.
If the value of $p$ equals 0, $(e')_{10}$ is still assigned with $0$.
Otherwise, the value of the new exponent $(e')_{10}$ is the difference between $(e)_{10}$ and $min$, representing the offset by the $min$.
Because $0$ already indicates a 0 value, $(e')_{10}$ must add $1$ to shift the offset by 1 and is then stored in $p'$.
The quantized $p'$ with $(e')_{10}$ as the new exponent is finally stored in the quantized model $\Lambda'$.
Taking the range [$\pm0$, $\pm2^{-29}$, \ldots, $\pm2^{0}$] for example, we have $max$ of $0$ and $min$ of $-29$.
Thus, we only need 6 ( = 1+$\lceil$ log$_{2}[0-(-29)+2] \rceil$) bits, whose 1 is the sign bit, to represent all of the possible values.
The quantized exponents $(e')_{10}$ of $\pm0$, $\pm2^{-29}$ and $\pm2^{0}$ are $0, 1$ and $30$ respectively.
It is clear that there is no performance degradation when applying the exponent-quantization since we only reduce the number of bits to represent a parameter value instead of changing the value.


%
\section{Experiments}\label{sec:Experiments}
This section presents the experimental results of the EOFP-QNN on the SE task. We used two standardized evaluation metrics: perceptual evaluation of speech quality (PESQ) \cite{PESQ2001}, and short-time objective intelligibility measure (STOI) \cite{STOI2011}, to test the performance. PESQ was designed to evaluate the quality of processed speech, and the score ranges from -0.5 to 4.5. A higher PESQ score denotes that the enhanced speech has better speech quality. STOI was designed to compute the speech intelligibility, and the score ranges from 0 to 1. A higher STOI value indicates better speech intelligibility.

\subsection{Experimental Setup}
The TIMIT corpus \cite{TIMIT1993} was used to prepare the training and test sets. For the training set, all of the 4620 training utterances from the TIMIT corpus were used and further corrupted with 100 different noise types at eight signal-to-noise (SNR) levels (from -10 dB to 25 dB with a step of 5 dB) in a random manner.  For the test set, we selected another 100 utterances (different from those used in the training set) from TIMIT and corrupted these utterances using another three noise signals (engine, street, and two talkers) at four SNR levels (-6, 0, 6, and 12 dB). Note that we intentionally designed both noise types and SNR levels of the training and test sets to be different to make the experimental conditions more realistic.

For the BLSTM spectral mapping system, the speech was parametrized into a sequence of 257-dimensional log-power spectral (LPS) features, and the mapping was performed in a frame-by-frame manner. The BLSTM model has two bidirectional LSTM layers, each with 257 nodes; one fully connected layer, with 300 nodes; and a fully connected output layer. We used the similar model structure as that used in \cite{x5}.
For the FCN waveform mapping system, the mapping was directly performed in the raw-waveform domain, and, thus, no additional analysis and restoration process were required. The FCM model used here shared the similar structure as that used in \cite{x6}, and an end-to-end utterance enhancement was carried out. The FCN model has ten convolutional layers with zero padding to preserve the same size as the input. The first ten layers consist of 30 filters, with a filter size of 55. There is one filter with a size of 55 in the last layer. In the experiments, we applied the proposed EOFP-QNN to both the BLSTM and the FCN models. For a fair comparison, the structure of the models and the number of the epochs for training were the same for the original full-precision model and the quantized model. 

\subsection{Experimental Result}

\subsubsection{Proposed method versus directly chopping in the mantissa-quantization}
As presented earlier, the EOFP technique applies a conditional rounding process to remove unnecessary bits in the mantissa-quantization. Another way to remove bits is directly chopping, namely, keeping the first $(32 - n)$ target bits and directly chopping the other $n$ bits. 
Here, we compared the performance of the conditional rounding and directly chopping processes for mantissa-quantization. 
We tested the performance using seven different bit-widths (bit-width = 26, 20, 14, 12, 11, 10 and 9) to compare with the non-quantized model (bit-width = 32) . 
The results are listed in Table~\ref{tab:exp2table}.
Each value in Table~\ref{tab:exp2table} is an average PESQ score over three noise types and four SNR levels. 
From the table, we first note that, when the proposed method was used, the PESQ suffered only marginal reductions. 
For example, when the bit number was reduced from 32 to 9, the PESQ reductions were 1.49\%  (2.144 to  2.112) and -0.58\% (from 2.064 to 2.076), respectively, for BLSTM and FCN, respectively. Note that the negative sign means that the performance even becomes better than the non-quantized model.
The results suggest that we may quantize all the mantissa bits and keep only 1 sign bit and 8 exponent bits to replace the original 32-bit data while maintaining similar enhancement performance. 
However, when we replaced our method with the directly chopping process, the PESQ scores were notably decreased. When reducing the bit number from 32 to 9 bits, we noted clear PESQ reductions of 2.15\% (from 2.144 to  2.098) and 9.88\% (from 2.064 to 1.860), respectively, for BLSTM and FCN models.

	\begin{table}[h]
		\caption{PESQ scores of quantized BLSTM and FCN using the proposed conditional rounding and directly chopping for mantissa-quantization with different bit-widths.} 
		\centering 
		\begin{tabular}{c c c c c c c } 
			\hline\hline 
			Bit- & \multicolumn{2}{c}{BLSTM} &  \multicolumn{2}{c}{FCN}\\ 
			width & Proposed & Chopping & Proposed & Chopping\\
			32 & 2.144 & 2.144 & 2.064 & 2.064 \\
            26 & 2.136 & 2.135 & 2.074 & 2.064 \\
            20 & 2.125 & 2.141 & 2.081 & 2.074 \\
            14 & 2.135 & 2.136 & 2.093 & 2.086 \\
            12 & 2.157 & 2.147 & 2.088 & 2.064 \\
            11 & 2.131 & 2.144 & 2.070 & 2.078 \\
            10 & 2.154 & 2.146 & 2.054 & 2.035 \\
            9 & 2.112 & 2.098 & 2.076 & 1.860 \\
			\hline 
		\end{tabular}
		\label{tab:exp2table}
	\end{table}
  

\subsubsection{BLSTM and FCN with mantissa-quantization}
From Table~\ref{tab:exp2table}, we can observe that the bit-width can be reduced from 32 to 9 with only a marginal PESQ performance drop for the BLSTM and FCN model. 
In Table~\ref{tab:exp1table}, we listed the detailed PESQ and STOI scores under specific SNR levels for both the BLSTM and the FCN system. Each value is an average score of three noise types. The scores of the unprocessed noisy speech are also shown for comparison. 

	\begin{table*}[t]
		\caption{Detailed PESQ and STOI scores for the original and quantized models under specific SNR conditions. The quantized models were quantized by the mantissa-quantization (with 9 bit-width).} 
		\centering 
		\label{tab:exp1table}
		\begin{tabular}{|c | cc | cc | cc | cc | cc |}
			\hline
			\multirow{2}{*}{~}&\multicolumn{2}{|c|}{\multirow{2}{*}{Noisy}}
			 & \multicolumn{4}{|c|}{BLSTM (LPS)} & \multicolumn{4}{|c|}{FCN (Raw waveform)}\\
			 \cline{4-11} 
			 &\multicolumn{2}{c}{} & \multicolumn{2}{|c|}{ Original} & \multicolumn{2}{|c|}{   Quantized} &
			 \multicolumn{2}{|c|}{   Original  } &
			 \multicolumn{2}{|c|}{   Quantized  } \\
			\hline
			
			\hline
			
			SNR(dB) & PESQ & STOI & PESQ & STOI & PESQ & STOI & PESQ & STOI & PESQ & STOI \\ 
			\hline 
			
			\hline
			-6dB & 1.223 & 0.509 & 1.499 & 0.568 & 1.488 & 0.569 & 1.381 & 0.548 & 1.444 & 0.538 \\
			\hline
			0dB & 1.622 & 0.659 & 1.983 & 0.728 & 1.962 & 0.725 & 1.843 & 0.719 & 1.877 & 0.700 \\
			\hline
			6dB & 2.016 & 0.800 & 2.393 & 0.832 & 2.361 & 0.831 & 2.304 & 0.840 & 2.281 & 0.814 \\
			\hline
			12dB & 2.439 & 0.901 & 2.699 & 0.885 & 2.638 & 0.884 & 2.729 & 0.911 & 2.700 & 0.878 \\
			\hline
			
			\hline
			Average & 1.825 & 0.717 & 2.144 & 0.753 & 2.112 & 0.752 & 2.064 & 0.755 & 2.076 & 0.733 \\
			
%
			\hline		
		\end{tabular}
	\end{table*}


From Table~\ref{tab:exp1table}, we first note that, when the EOFP quantization technique was applied, there was only a 1.49\% (from 2.144 to 2.112) PESQ score reduction and a 0.13\% (from 0.753 to 0.752) STOI score reduction for the BLSTM system. Similarly for the FCN system, we note a -0.54\% (2.064 to 2.076) PESQ reduction and a 2.91\% (from 0.755 to 0.733) STOI reduction. Note that in this set of experiments, we quantized every parameter in the model from a 32-bit floating point to a 9-bit exponent. The total compression ratio was 3.56. The results in Table 3 confirm that, although the model size had been notably compressed, the objective quality and intelligibility scores were only marginally reduced. We also noted that FCN suffered more STOI reductions than BLSTM after quantization. 
A possible reason is that FCN includes  comparatively fewer parameters than BLSTM. Therefore, each parameter in FCN plays a more important role than BLSTM, and thus model quantization induces a bit stronger influence. 

\subsubsection{BLSTM and FCN with exponent-quantization}	
\begin{figure}[ht]
		\centering
		\includegraphics[width=1\columnwidth]{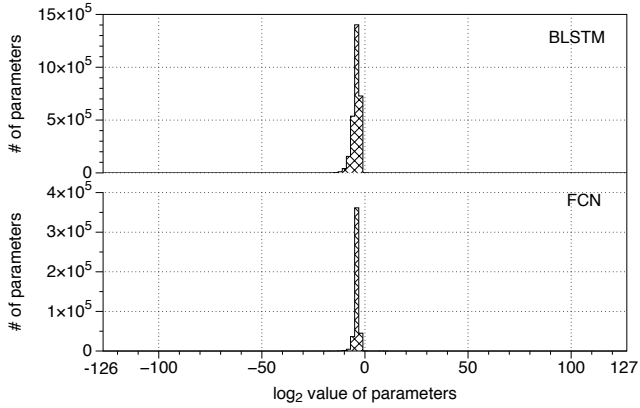}
		\caption{Distribution of the model log$_{2}$ parameter values for BLSTM (top) and FCN (bottom). }\label{fig:exp1distri}
	\end{figure}

Next, we apply the exponent-quantization to further reduce the model size. The overall quantization is termed "mantissa+exponent-quantization" in the following discussion. As mentioned in Section 3.3, we first need to identify the optimal bit-width before quantization. To this end, we examined the distribution of the log$_{2}$ value of all the parameters in BLSTM and FCN. 
The results are shown in Fig.~\ref{fig:exp1distri}. From the figure, most parameters in the two models are distributed in a narrow region, suggesting that we are allowed to further reduce the bit-width. Next, we calculated the maximum and the minimum log$_{2}$ value of each model and the bit-width from Algorithm 2. Then, we obtained the \{$max$, $min$, $len$\} as \{0, -23, 5\} and as \{10, -26, 6\} for the BLSTM and FCN models, respectively. 
On the basis of the computed \{$max$, $min$, $len$\}, we can further perform exponent-quantization on the BLSTM and FCN models. 
The quantization results of the two models using the {mantissa-quantization}, and the {mantissa+exponent-quantization} are listed in the fourth and fifth raws, respectively, in Table~\ref{tab:exp1posttable}. 
	
	\begin{table}[h]
		\caption{Number of parameters and the corresponding bytes used in the BLSTM and FCN before and after quantization. } 
		\centering 
		\begin{tabular}{l c c } 
			\hline\hline 
			~ & BLSTM & FCN \\
			Number of parameters & 2,877,929& 450,301  \\
			Single-precision size (KB) & 11,242 & 1,759 \\
			Mantissa-quantization (KB) & 3,162 & 495 \\
			+ Exponent-quantization (KB) & 2,108 & 385 \\ 
			\hline 
		\end{tabular}
		\label{tab:exp1posttable}
	\end{table}
	
	
From Tables~\ref{tab:exp1table} and~\ref{tab:exp1posttable} , we can see that the model sizes of the "mantissa-quantization" and "mantissa+exponent-quantization" quantized
BLSTM models were only 28.13\% (3,162/11,242) and 18.75\% (2,108/11,242), respectively, when compared to the original (non-quantized) BLSTM model. Please note that the exponent-quantization only further reduced the model size but did not cause extra PESQ and STOI reductions. We observed similar trends for the FCN models. The model sizes of the "mantissa-quantization" and "mantissa+exponent-quantization" quantized FCN models were only 28.14\% (495/1,759) and 21.89\% (385/1,759), respectively, when comparing to the original (non-quantized) FCN model.
On the basis of the above observations, we can conclude that, by using the proposed EOFP-QNN (mantissa+exponent-quantization), we can significantly reduce the model sizes of BLSTM and FCN while maintaining satisfactory quality and intelligibility scores as compared to the original non-quantized models. 

\section{Conclusions}\label{sec:conclusion}
In this work, we proposed a novel EOFP-QNN and evaluated its effect on SE performance. To the best of our knowledge,
this is the first study that investigates the effect of model compression based on the floating-point quantization technique on the SE task.
The results showed that, by applying the EOFP, the model sizes of the quantized models were only 18.75\% and 21.89\% for BLSTM and FCN, respectively, compared to the original models.
With such significant model size reductions, the quality and intelligibility scores were only marginally degraded. For example, the PESQ and STOI score reductions were 1.49\% and 0.13\% for the BLSTM SE system. The results suggest that, by using the proposed EOFP quantization technique, we may be able to install an SE system with a compressed DL-based model in embedded devices to operate in an IoT environment.

\bibliographystyle{IEEEbib}
\bibliography{refs}

\begin{thebibliography}{10}

\bibitem{imgRec1}
K.~Simonyan and A.~Zisserman,
\newblock ``{Very deep convolutional networks for large-scale image
  recognition},''
\newblock in {\em Proc. ICLR}, 2015.

\bibitem{ResNet2016}
K.~He, X.~Zhang, S.~Ren, and J.~Sun,
\newblock ``Deep residual learning for image recognition,''
\newblock in {\em Proc. CVPR}, 2016, pp. 770--778.

\bibitem{Speech2013}
A.~Graves, A.~Mohamed, and G.~Hinton,
\newblock ``Speech recognition with deep recurrent neural networks,''
\newblock in {\em Proc. ICASSP}, 2013, pp. 6645--6649.

\bibitem{Speech2012}
G.~Hinton, L.~Deng, D.~Yu, G.~Dahl, A.~Mohamed, N.~Jaitly, A.~Senior,
  V.~Vanhoucke, P.~Nguyen, T.~Sainath, et~al.,
\newblock ``Deep neural networks for acoustic modeling in speech recognition:
  The shared views of four research groups,''
\newblock {\em IEEE Signal processing magazine}, vol. 29, no. 6, pp. 82--97,
  2012.

\bibitem{Jinyu2013}
L.~Deng, J.~Li, J.-T. Huang, K.~Yao, D.~Yu, F.~Seide, M.~Seltzer, G.~Zweig,
  X.~He, J.~Williams, et~al.,
\newblock ``Recent advances in deep learning for speech research at
  microsoft,''
\newblock in {\em Proc. ICASSP}, 2013, pp. 8604--8608.

\bibitem{obj1}
P.~Luo, Y.~Tian, X.~Wang, and X.~Tang,
\newblock ``Switchable deep network for pedestrian detection,''
\newblock in {\em Proc. CVPR}, 2014, pp. 899--906.

\bibitem{obj2}
X.~Zeng, W.~Ouyang, and X.~Wang,
\newblock ``Multi-stage contextual deep learning for pedestrian detection,''
\newblock in {\em Proc. ICCV}, 2013, pp. 121--128.

\bibitem{obj3}
P.~Sermanet, K.~Kavukcuoglu, S.~Chintala, and Y.~LeCun,
\newblock ``Pedestrian detection with unsupervised multi-stage feature
  learning,''
\newblock in {\em Proc. CVPR}, 2013, pp. 3626--3633.

\bibitem{refNLU}
R.~Collobert and J.~Weston,
\newblock ``A unified architecture for natural language processing: Deep neural
  networks with multitask learning,''
\newblock in {\em Proc. ICML}, 2008, pp. 160--167.

\bibitem{Mikolov2013}
T.~Mikolov, I.~Sutskever, K.~Chen, G.~Corrado, and J.~Dean,
\newblock ``Distributed representations of words and phrases and their
  compositionality,''
\newblock in {\em Proc. NIPS}, 2013, pp. 3111--3119.

\bibitem{Mikolov2012}
T.~Mikolov, M.~Karafi{\'a}t, L.~Burget, J.~{\v{C}}ernock{\`y}, and
  S.~Khudanpur,
\newblock ``Recurrent neural network based language model,''
\newblock in {\em Proc. Interspeech}, 2010, pp. 1045--1048.

\bibitem{refTPU}
N.~P. Jouppi, C.~Young, N.~Patil, D.~Patterson, G.~Agrawal, R.~Bajwa, S.~Bates,
  S.~Bhatia, N.~Boden, A.~Borchers, et~al.,
\newblock ``{In-datacenter performance analysis of a tensor processing unit},''
\newblock in {\em Proc. ISCA}, 2017, pp. 1--12.

\bibitem{refIoT}
Y.~K. Chen,
\newblock ``Challenges and opportunities of internet of things,''
\newblock in {\em Proc. ASPDAC}, 2012, pp. 383--388.

\bibitem{BinaryConnect}
M.~Courbariaux, Y.~Bengio, and J.-P. David,
\newblock ``Binaryconnect: Training deep neural networks with binary weights
  during propagations,''
\newblock in {\em Proc. NIPS}, 2015, pp. 3105--3113.

\bibitem{G2014}
Y.~Gong, L.~Liu, M.~Yang, and L.~Bourdev,
\newblock ``Compressing deep convolutional networks using vector
  quantization,''
\newblock {\em arXiv preprint arXiv:1412.6115}, 2014.

\bibitem{INQ}
A.~Zhou, A.~Yao, Y.~Guo, L.~Xu, and Y.~Chen,
\newblock ``Incremental network quantization: Towards lossless cnns with
  low-precision weights,''
\newblock in {\em Proc. ICLR}, 2017.

\bibitem{bridgeC}
P.~H. Hung, C.~H. Lee, S.~W. Yang, V.~S. Somayazulu, Y.~K. Chen, and S.~Y.
  Chien,
\newblock ``Bridge deep learning to the physical world: An efficient method to
  quantize network,''
\newblock in {\em Proc. SiPS}, 2015, pp. 1--6.

\bibitem{qspeech1}
K.~Hwang and W.~Sung,
\newblock ``Fixed-point feedforward deep neural network design using weights
  +1, 0, and −1,''
\newblock in {\em Proc. SiPS}, 2014, pp. 1--6.

\bibitem{qspeech2}
F.~Seide, H.~Fu, J.~Droppo, G.~Li, and D.~Yu,
\newblock ``1-bit stochastic gradient descent and its application to
  data-parallel distributed training of speech dnns,''
\newblock in {\em Proc. Interspeech}, 2014, pp. 1058--1062.

\bibitem{qspeech3}
R.~Prabhavalkar, O.~Alsharif, A.~Bruguier, and L.~McGraw,
\newblock ``On the compression of recurrent neural networks with an application
  to lvcsr acoustic modeling for embedded speech recognition,''
\newblock in {\em Proc. ICASSP}, 2016, pp. 5970--5974.

\bibitem{qspeech4}
S.~Han, J.~Kang, H.~Mao, Y.~Hu, X.~Li, Y.~Li, D.~Xie, H.~Luo, S.~Yao, Y.~Wang,
  H.~Yang, and W.~J. Dally,
\newblock ``Ese: Efficient speech recognition engine with sparse lstm on
  fpga,''
\newblock in {\em Proc. FPGA}, 2017, pp. 75--84.

\bibitem{qspeech5}
Y.~Lin, S.~Han, H.~Mao, Y.~Wang, and W.~Dally,
\newblock ``Deep gradient compression: Reducing the communication bandwidth for
  distributed training,''
\newblock in {\em Proc. ICLR}, 2018.

\bibitem{qspeech6}
Y.~Wang, J.~Li, and Y.~Gong,
\newblock ``Small-footprint high-performance deep neural network-based speech
  recognition using split-vq,''
\newblock in {\em Proc. ICASSP}, 2015, pp. 4984--4988.

\bibitem{ko2017precision}
J.~H. Ko, J.~Fromm, M.~Philipose, I.~Tashev, and S.~Zarar,
\newblock ``Precision scaling of neural networks for efficient audio
  processing,''
\newblock {\em arXiv preprint arXiv:1712.01340}, 2017.

\bibitem{x1}
D.~Wang and J.~Chen,
\newblock ``Supervised speech separation based on deep learning: An overview,''
\newblock {\em IEEE/ACM Transactions on Audio, Speech, and Language
  Processing}, vol. 26, no. 10, pp. 1702--1726, Oct 2018.

\bibitem{x2}
X.~Lu, Y.~Tsao, S.~Matsuda, and C.~Hori,
\newblock ``Speech enhancement based on deep denoising autoencoder,''
\newblock in {\em Proc. Interspeech}, 2013, pp. 436--440.

\bibitem{xia2014wiener}
B.~Xia and C.~Bao,
\newblock ``Wiener filtering based speech enhancement with weighted denoising
  auto-encoder and noise classification,''
\newblock {\em Speech Communication}, vol. 60, pp. 13--29, 2014.

\bibitem{x3}
Y.~Xu, J.~Du, L.~R. Dai, and C.~H. Lee,
\newblock ``A regression approach to speech enhancement based on deep neural
  networks,''
\newblock {\em IEEE/ACM Transactions on Audio, Speech, and Language
  Processing}, vol. 23, no. 1, pp. 7--19, Jan 2015.

\bibitem{kolbk2017speech}
M.~Kolbk, Z.-H. Tan, J.~Jensen, M.~Kolbk, Z.-H. Tan, and J.~Jensen,
\newblock ``Speech intelligibility potential of general and specialized deep
  neural network based speech enhancement systems,''
\newblock {\em IEEE/ACM Transactions on Audio, Speech, and Language
  Processing}, vol. 25, no. 1, pp. 153--167, Nov 2017.

\bibitem{x4}
S.-W. Fu, Y.~Tsao, and X.~Lu,
\newblock ``S{N}{R}-aware convolutional neural network modeling for speech
  enhancement,''
\newblock in {\em Proc. Interspeech}, 2016, pp. 3768--3772.

\bibitem{fu2017complex}
S.-W. Fu, T.-Y Hu, Y.~Tsao, and X.~Lu,
\newblock ``Complex spectrogram enhancement by convolutional neural network
  with multi-metrics learning,''
\newblock in {\em MLSP}, 2017, pp. 1--6.

\bibitem{x5}
H.~Erdogan, S.~Watanabe J.~R.~Hershey, and J.~Le Roux,
\newblock ``Phase-sensitive and recognition-boosted speech separation using
  deep recurrent neural networks,''
\newblock in {\em Proc. ICASSP}, 2015, pp. 708--712.

\bibitem{chen2015speech}
Z.~Chen, S.~Watanabe, H.~Erdogan, and J.~R. Hershey,
\newblock ``Speech enhancement and recognition using multi-task learning of
  long short-term memory recurrent neural networks,''
\newblock in {\em Proc. Interspeech}, 2015, pp. 1--5.

\bibitem{fu2017raw}
S.-W. Fu, Y.~Tsao, X.~Lu, and H.~Kawai,
\newblock ``Raw waveform-based speech enhancement by fully convolutional
  networks,''
\newblock {\em arXiv preprint arXiv:1703.02205}, 2017.

\bibitem{x6}
S.-W. Fu, T.-W. Wang, Y.~Tsao, X.~Lu, and H.~Kawai,
\newblock ``End-to-end waveform utterance enhancement for direct evaluation
  metrics optimization by fully convolutional neural networks,''
\newblock {\em IEEE/ACM Transactions on Audio, Speech, and Language
  Processing}, vol. 26, no. 9, pp. 1570--1584, Sept 2018.

\bibitem{x7}
S.~Pascual, A.~Bonafonte, and J.~Serrà,
\newblock ``Segan: Speech enhancement generative adversarial network,''
\newblock in {\em Proc. Interspeech}, 2017, pp. 3642--3646.

\bibitem{x8}
A.~Van Den~Oord, S.~Dieleman, H.~Zen, K.~Simonyan, O.~Vinyals, A.~Graves,
  N.~Kalchbrenner, A.~Senior, and K.~Kavukcuoglu,
\newblock ``Wavenet: A generative model for raw audio,''
\newblock {\em arXiv preprint arXiv:1609.03499}, 2016.

\bibitem{sun2017optimization}
H.~Sun and S.~Li,
\newblock ``An optimization method for speech enhancement based on deep neural
  network,''
\newblock in {\em IOP Conference Series: Earth and Environmental Science}. IOP
  Publishing, 2017, vol.~69, p. 012139.

\bibitem{IEEE754}
Institute of~Electrical and Electronics Engineers,
\newblock ``Ieee standard for binary floating-point arithmetic,''
\newblock {\em ANSI/IEEE Std 754-1985}, 1985.

\bibitem{PESQ2001}
ITU-T Recommendation,
\newblock ``Perceptual evaluation of speech quality (pesq): An objective method
  for end-to-end speech quality assessment of narrow-band telephone networks
  and speech codecs,''
\newblock {\em Rec. ITU-T P. 862}, Jan 2001.

\bibitem{STOI2011}
C.~H. Taal, R.~C. Hendriks, R.~Heusdens, and J.~Jensen,
\newblock ``An algorithm for intelligibility prediction of time--frequency
  weighted noisy speech,''
\newblock {\em IEEE Transactions on Audio, Speech, and Language Processing},
  vol. 19, no. 7, pp. 2125--2136, Sept 2011.

\bibitem{TIMIT1993}
J.~S. Garofolo, L.~Lamel, W.~M. Fisher, J.~G. Fiscus, and D.~S. Pallett,
\newblock ``Darpa timit acoustic-phonetic continous speech corpus cd-rom. nist
  speech disc 1-1.1,''
\newblock {\em NASA STI/Recon technical report n}, vol. 93, 1993.

\end{thebibliography}

\end{document}